\documentclass[aps,prl,showpacs,twocolumn,amsmath,amssymb]{revtex4}
\usepackage[english]{babel}
\usepackage{latexsym}
\usepackage{graphics}
\usepackage{subfigure}
\usepackage{epsfig}

\begin{document}

\title{Suppression of Nonlinear Interactions in Resonant Macroscopic Quantum Devices~:\\
the Example of the Solid-State Ring Laser Gyroscope}

\author{Sylvain~Schwartz$^{1}$, Fran\c{c}ois Gutty$^{2}$, Gilles~Feugnet$^1$, Philippe~Bouyer$^3$
and Jean-Paul~Pocholle$^1$}

\affiliation{$^1$Thales Research and Technology France, Route D\'epartementale 128, F-91767 Palaiseau Cedex, France\\
$^2$Thales Avionics, 40 rue de la Brelandi\`ere, BP 128, F-86101 Ch\^{a}tellerault, France \\ $^3$Laboratoire Charles
Fabry de l'Institut d'Optique, CNRS and Universit\'e Paris-Sud, Campus Polytechnique, RD 128, F-91127 Palaiseau Cedex,
France}
\date{\today}

\pacs{42.65.Sf, 42.62.Eh, 06.30.Gv, 42.55.Rz}

\email{sylvain.schwartz@thalesgroup.com}

\begin{abstract}
We study the suppression of nonlinear interactions in resonant macroscopic quantum devices in the case of the
solid-state ring laser gyroscope. These nonlinear interactions are tuned by vibrating the gain medium along the cavity
axis. Beat note occurrence under rotation provides a precise measurement of the strength of nonlinear interactions,
which turn out to vanish for some discrete values of the amplitude of vibration. Our theoretical description, in very
good agreement with the measured data, suggests the use of a higher vibration frequency to achieve quasi-ideal rotation
sensing over a broad range of rotation speeds. We finally underline the analogy between this device and some other
macroscopic quantum rotation sensors, such as ring-shaped superfluid configurations, where nonlinear interactions could
be tuned for example by the use of magnetically-induced Feschbach resonance.
\end{abstract}

\maketitle

The use of macroscopic quantum effects for rotation sensing in ring-shaped configurations has been extensively studied
in the case of both optical systems \cite{Aronowitz, khanin, russes} and superfluids \cite{Bloch, Leggett1}, either
liquid helium \cite{Varoquaux,Packard} or Bose-Einstein condensed gases \cite{Mueller, Benakli, Jackson, Schwartz_BEC,
Phillips}. As pointed out in \cite{Leggett2, Schwartz_PRL}, nonlinear interactions play a crucial role in the dynamics
of such devices, as they can hinder or affect their ability to sense rotation, even when counteracted by other coupling
sources. Consequently, the possibility of tuning or even suppressing nonlinear interactions is of great importance for
using these devices as rotation sensors.

Several systems offer the possibility of controlling the strength of their nonlinearities. For example, in the case of
gas ring laser gyroscopes, one can considerably lower mode competition by tuning the cavity out of resonance with the
atoms at rest, resulting in the quasi-suppression of nonlinear interactions \cite{Aronowitz}. In the case of atomic
systems, it is also possible to tune and even suppress nonlinear interactions, by using Feshbach resonance
\cite{Nature98, PRL98, inguscio}. As regards solid-state ring lasers, we have recently demonstrated \cite{Schwartz_PRL}
the possibility of stable rotation sensing thanks to the circumvention of mode competition by the use of an additional
stabilizing coupling. However, nonlinear interactions are still present in this configuration, and can even be
quantitatively observed \cite{Schwartz_PRL,Schwartz2}.

In this Letter, we report the experimental and theoretical study of a novel technique intended to tune and suppress
nonlinear interactions in a solid-state ring laser gyroscope, similarly to the case of scattering length control in an
atomic system. This is achieved by vibrating the gain crystal along the optical axis of the laser cavity, considering
the fact that nonlinear interactions in a solid-state ring laser result mainly from mutual coupling between the
counterpropagating modes induced by the population inversion grating established in the amplifying medium \cite{russes,
Schwartz_PRL}. Using the quantitative information on the strength of the nonlinear interactions provided by the beat
note between the counterpropagating laser beams \cite{Schwartz_PRL}, we demonstrate experimentally the possibility of
suppressing these interactions for some discrete values of the amplitude of the crystal movement. We eventually derive,
in the limit of high vibration frequencies, a very simple condition for rotation sensing and point out the similarity
with the equivalent condition for a toroidal Bose-Einstein condensed gas, resulting from the toy model of
\cite{Leggett2} where the effects of scattering length tuning described in \cite{Nature98} are included.

The solid-state ring laser gyroscope can be described semiclassically, assuming one single identical mode in each
direction of propagation (something which is guaranteed by the attenuation of spatial hole burning effects thanks to the
gain crystal movement \cite{danielmeyer}), one single identical state of polarization and plane wave approximation. The
electrical field inside the cavity can then be written as follows~:
\begin{equation}
E(x,t) = \textrm{Re} \left\{ \sum_{p=1}^2 \tilde{E}_{p} (t) e^{i(\omega_c t + \mu_{p} kx)} \right\} \;, \nonumber
\end{equation}
where $\mu_p = (-1)^p$ and where $\omega_c$ and $k$ are respectively the angular and spatial average frequencies of the
laser, whose longitudinal axe is associated with the $x$ coordinate. In the absence of crystal vibration, the equations
of evolution for the slowly-varying amplitudes $\tilde{E}_{1,2}$ and for the population inversion density $N$ have the
following expression \cite{russes,Schwartz_PRL}~:
\begin{eqnarray}
\frac{\textrm{d}\tilde{E}_{1,2}}{\textrm{d}t}& = & -\frac{\gamma_{1,2}}{2}
\tilde{E}_{1,2}+i\frac{\tilde{m}_{1,2}}{2}\tilde{E}_{2,1} + i \mu_{1,2} \frac{\Omega}{2}\tilde{E}_{1,2} \label{MaxBloch}\\
 & & \!\!\!\!\!\!\!\!\!\!\!\!\!\!\!\!\!\!\!\!\!\!\!\!+\frac{\sigma}{2T} \left( \tilde{E}_{1,2}
\int_0^L N \textrm{d}x+ \tilde{E}_{2,1}\int_0^L N{e}^{-2i \mu_{1,2} kx}\textrm{d} x \right) \;, \nonumber
\end{eqnarray}
\begin{equation}
\!\!\!\!\!\!\!\!\!\!\!\!\!\!\!\!\!\!\!\!\!\!\!
  \frac{\partial N}{\partial t} = W_\textrm{th}(1+\eta)-\frac{N}{T_1}-\frac{aN E(x,t)^2 }{T_1} \;, \label{MaxBloch2}
\end{equation}
where $\gamma_{1,2}$ are the intensity losses per time unit for each mode, $\tilde{m}_{1,2}$ are the backscattering
coefficients, $\Omega$ is the difference between the eigenfrequencies of the counterpropagating modes (including the
effect of rotation, see further), $\sigma$ is the laser cross section, $T$ is the cavity round-trip time, $\eta$ is the
relative excess of pumping power above the threshold value $W_\textrm{th}$, $T_1$ is the lifetime of the population
inversion and $a$ is the saturation parameter. Throughout this paper we shall neglect dispersion effects, considering
the fact that the Nd-YAG gain width is much larger than the laser cavity free spectral range. The backscattering
coefficients, which depend on spatial inhomogeneities of the propagation medium \cite{spreeuw}, have the following
expression \cite{Schwartz2}~:
\begin{equation} \label{back}
\tilde{m}_{1,2}= -\frac{\omega_c}{\bar{\varepsilon} cT} \oint_0^L \left[ \varepsilon(x) - \frac{i \kappa(x)}{\omega_c}
\right] e^{-2i \mu_{1,2} k x} \textrm{d}x \; ,
\end{equation}
where $\varepsilon (x)$ and $\kappa(x)$ are respectively the dielectric constant and the fictitious conductivity along
the cavity perimeter in the framework of an ohmic losses model \cite{Siegman}, where $c$ is the speed of light in vacuum
and where $\bar{\varepsilon}$ stands for the spatial average of $\varepsilon$. In order to counteract mode competition
effects and ensure beat regime operation under rotation, an additional stabilizing coupling as described in
\cite{Schwartz_PRL} is introduced, resulting in losses of the following form~:
\begin{equation} \label{gamma}
\gamma_{1,2} = \gamma - \mu_{1,2} K a (|\tilde{E}_1|^2-|\tilde{E}_2|^2) \;,
\end{equation}
where $\gamma=\bar{\kappa}/\bar{\varepsilon}$ is the average loss coefficient and where $K>0$ represents the strength of
the stabilizing coupling.

\begin{figure}
\begin{center}
\includegraphics[scale=0.6]{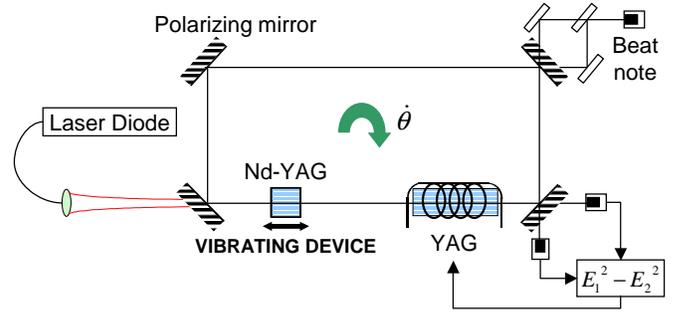}
\end{center}
\caption{Scheme of our experimental setup. The diode-pumped vibrating Nd-YAG crystal is placed inside a 22-cm ring
cavity on a turntable. Losses of the form (\ref{gamma}) are created by a feedback loop acting on a Faraday rotator (an
additional YAG crystal inside a solenoid), in combination with a polarizing mirror and a slight non-planarity of the
cavity (not drawn here). Two photodiodes are used for generating the error signal of the feedback loop. A third
photodiode measures the frequency of the beat note between the counterpropagating modes.}\label{figure 1}
\end{figure}

We assume the following sinusoidal law to account for the gain crystal vibration~:
\begin{equation} \label{xc}
x_c(t) = \frac{x_m}{2} \sin(2\pi f_m t) \;,
\end{equation}
where $x_c(t)$ is the coordinate, in the frame of the laser cavity, of a given reference point attached to the crystal,
and where $x_m$ and $f_m$ are respectively the amplitude and the frequency of the vibration movement. The population
inversion density function in the frame of the vibrating crystal $N_c(x,t)$ is ruled by the following equation~:
\begin{equation} \label{Nc}
\frac{\partial N_c}{\partial t} = W_\textrm{th} (1+\eta) - \frac{N_c}{T_1} - \frac{aN_c E(x+x_c(t),t)^2}{T_1} \;,
\end{equation}
where $E(x,t)$ refers to the electric field in the cavity (non-vibrating) frame. Moreover, $N_c(x,t)$ can be deduced
from its equivalent in the cavity frame $N(x,t)$ by the identity $N_c(x,t) = N(x+x_c(t),t)$, resulting in the following
expressions~:
\begin{equation} \nonumber
\left\{
\begin{split}
&\int_0^L N(x,t) \textrm{d}x = \int_0^L N_c(x,t) \textrm{d}x \;, \\
&\int_0^L N(x,t) e^{2ikx} \textrm{d}x = e^{2ikx_c(t)} \int_0^L N_c(x,t) e^{2ikx} \textrm{d}x \;.
\end{split}
\right.
\end{equation}
The backscattering coefficients (\ref{back}) acquire in the presence of the crystal vibration the following
time-dependent form~:
\begin{equation} \label{back2}
\tilde{m}_{1,2}(t)= \tilde{m}_{1,2}^{\phantom{c}c} e^{-2i \mu_1,2 k x_c(t)} + \tilde{m}_{1,2}^{\phantom{c}m} \; ,
\end{equation}
where $\tilde{m}_{1,2}^{\phantom{c}c}$ and $\tilde{m}_{1,2}^{\phantom{c}m}$, which are time-independent, account for the
backscattering due respectively to the crystal at rest and to any other diffusion source inside the laser cavity
(including the mirrors). As regards the difference $\Omega$ between the eigenfrequencies of the counterpropagating
modes, it results from the combined effects of the rotation (Sagnac effect \cite{Sagnac}) and of the crystal movement in
the cavity frame (Fresnel-Fizeau drag effect \cite{Fizeau}), resulting in the following expression~:
\begin{equation} \label{omega}
\frac{\Omega}{2\pi} = \frac{4A}{\lambda L} \dot{\theta} - \frac{2 \dot{x}_c(t)l(n^2-1)}{\lambda L} \;,
\end{equation}
where $A$ is the area enclosed by the ring cavity, $\lambda=2\pi c /\omega_c$ is the emission wavelength, $\dot{\theta}$
is the angular velocity of the cavity around its axis, and $l$ and $n$ are respectively the length and the refractive
index of the crystal (dispersion terms are shown to be negligible in this case).

The dynamics of the solid-state ring laser gyroscope with a vibrating gain medium is eventually ruled, in the framework
of our theoretical description, by the following equations~:
\begin{eqnarray}
\frac{\textrm{d}\tilde{E}_{1,2}}{\textrm{d}t}& = & -\frac{\gamma_{1,2}}{2}
\tilde{E}_{1,2}+i\frac{\tilde{m}_{1,2}}{2}\tilde{E}_{2,1} + i \mu_{1,2} \frac{\Omega}{2}\tilde{E}_{1,2} \label{MaxBlochvib}\\
 & & \!\!\!\!\!\!\!\!\!\!\!\!\!\!\!\!\!\!\!\!\!\!\!\!\!\!\!+\frac{\sigma}{2T} \left( \tilde{E}_{1,2}
\int_0^L N_c \textrm{d}x+ \tilde{E}_{2,1}e^{2ikx_c}\int_0^L N_c{e}^{-2i \mu_{1,2} kx}\textrm{d} x \right) \;, \nonumber
\end{eqnarray}
where $\gamma_{1,2}$, $x_c$, $N_c$, $\tilde{m}_{1,2}$ and $\Omega$ obey respectively equations (\ref{gamma}),
(\ref{xc}), (\ref{Nc}), (\ref{back2}) and (\ref{omega}). It comes out from this analysis that the solid-state ring laser
benefits, as a rotation sensor, from the crystal vibration in three separate and complementary ways~:
\begin{itemize}
\item the contrast of the population inversion grating, which is responsible for nonlinear coupling, is reduced on both
conditions that the amplitude of the movement is of the same order of magnitude than the step of the optical grating
(typically a fraction of $\mu$m) and that the period of the movement $1/f_m$ is significantly larger than the population
inversion response time $T_1$; the atomic dipoles are then no longer confined into a nodal or an antinodal area --~see
eq.~(\ref{Nc})~--, and become sensitive to the time-average value of the electric field, which can be independent of
their position on the crystal (at least when the laser is not rotating) provided the condition $J_0 (kx_m) = 0$ is
obeyed \cite{ulrich}, $J_0$ referring to the zero-order Bessel's function);
\item the light backscattered on the gain crystal from one mode into the other can be shifted out of resonance by the
Doppler effect resulting from the crystal movement in the cavity frame; this phenomenon, which induces a decrease of the
corresponding coupling strength, has previously been reported in the case of vibrating mirrors \cite{SPIE,diels}; in our
model, it arises from the time-dependent phase factor $\exp(2ikx_c)$ in front of the coupling coefficients $\tilde
m_{1,2}^{\phantom{c}c}$ and $\int N_c \textrm{d}x$ in equations (\ref{back2}) and (\ref{MaxBlochvib});
\item the frequency non-reciprocity between the counterpropagating modes due to the Fresnel-Fizeau dragging
effect --~eq.~(\ref{omega})~-- has a similar role as the mechanical dithering typically used for circumventing the
lock-in problem in the case of usual gas ring laser gyroscopes \cite{Chow}.
\end{itemize}

\begin{figure}
\begin{center}
\includegraphics[scale=0.9]{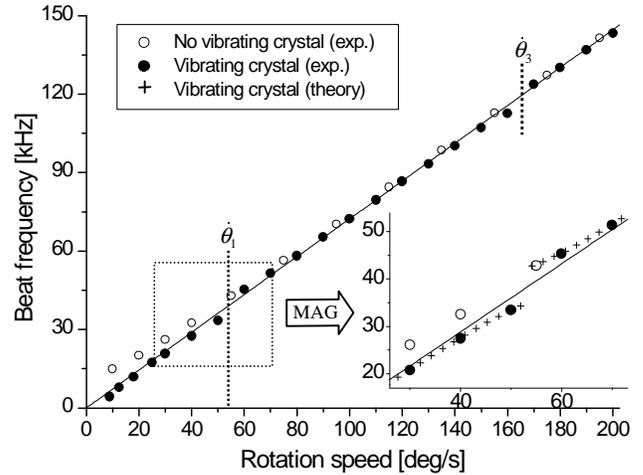}
\end{center}
\caption{Experimental beat frequency as a function of the rotation speed. White and black circles refer respectively to
the situations where the crystal is at rest and where the crystal is vibrating with a frequency $f_m\simeq 40$~kHz and
an amplitude $x_m \simeq 0.74$~$\mu$m. The insert shows a magnification around $\dot{\theta}_1$ --~see
eq.~(\ref{param})~--, together with theoretical predictions resulting from numerical simulations with the following
measured \cite{Schwartz2} parameters~: $\gamma=15.34$~$10^6$~s$^{-1}$, $\eta=0.21$, $|\tilde{m}_{1,2}^{\;c}|=
1.5$~$10^4$~s$^{-1}$, $|\tilde{m}_{1,2}^{\;m}|= 8.5$~$10^4$~s$^{-1}$, $\arg(\tilde{m}_{1}^{\;c}/\tilde{m}_{2}^{\;c})=
\arg(\tilde{m}_{1}^{\;m}/\tilde{m}_{2}^{\;m}) = \pi/17$, $K=10^7$~s$^{-1}$. Integration step is 0.1~$\mu$s, average
values are computed between 8 and 10~ms.} \label{figure 2}
\end{figure}

The solid-state ring laser setup we used in our experiment is sketched on Fig.~\ref{figure 1}. Thanks to the additional
stabilizing coupling (\ref{gamma}), a beat note signal is observed above a critical rotation speed, whose frequency is
plotted on Fig.~\ref{figure 2}. It can be seen on this figure that the difference between the ideal Sagnac line and the
experimental beat frequency, which is a direct measurement of the nonlinear interactions \cite{Schwartz_PRL}, is
considerably reduced in the zone ranging from 10 to 40~deg/s. Some nonlinearities are observed around the discrete
values $\dot{\theta} \simeq 55$~deg/s and $\dot{\theta} \simeq 165$~deg/s, in agreement with our theoretical model. As a
matter of fact, analytical calculations starting from equation (\ref{MaxBlochvib}) reveal the existence of disrupted
zones centered on discrete values of the rotation speed $\dot{\theta}_q$ obeying the following equation~:
\begin{equation}\label{param}
\frac{4A}{\lambda L} \dot{\theta}_q = q f_m \quad \textrm{where $q$ is an integer} \;,
\end{equation}
the size of each disrupted zone being proportional to $J_q (kx_m)$. With our experimental parameters, the first critical
velocity corresponds to $\dot{\theta}_1=55.5$~deg/s, the zones observed on Fig.~\ref{figure 2} corresponding to the
cases $q=1$ and $q=3$. The numerical simulations shown on the insert of this figure are in good agreement with our
analytical and experimental data. Such a phenomenon of disrupted zones has been reported previously in the case of gas
ring laser gyroscopes with mechanical dithering. It is sometimes designed as `Shapiro steps' \cite{Chow}, in reference
to an equivalent effect in the field of Josephson junctions \cite{Shapiro}.

\begin{figure}
\begin{center}
\includegraphics[scale=0.9]{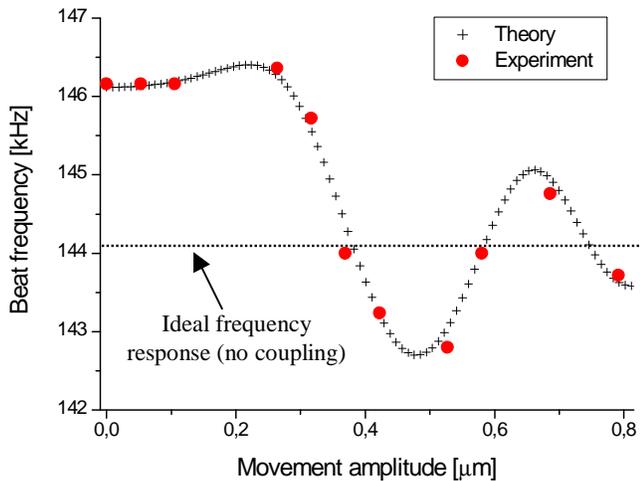}
\end{center}
\caption{Beat frequency as a function of the amplitude of the crystal movement for $\dot{\theta}=200$~deg/s. The
theoretical values (crosses) come from the numerical integration of equation (\ref{MaxBlochvib}) with the same
parameters as Fig.~\ref{figure 2}.}\label{figure 3}
\end{figure}

The dependence of the beat frequency on the amplitude of the crystal movement is shown on Fig.~\ref{figure 3}, for a
fixed rotation speed (200~deg/s). This graph illustrates the good agreement between our numerical simulations and our
experimental data. Moreover, this is an experimental demonstration of the direct control of the strength of nonlinear
interactions in the solid-state ring laser. In particular, for some special amplitudes of the crystal movement, the
influence of mode coupling vanishes, resulting in a beat frequency equal to the ideal Sagnac value.

This study suggests the use of a higher vibration frequency of the crystal, in order to increase the value of
$\dot{\theta}_1$ as much as possible. When $f_m \gg |\Omega|/(2\pi)$, the strength of the nonlinear interactions is
shown to be directly proportional to $J_0(k x_m)^2$, and the condition for rotation sensing reads~:
\begin{equation} \label{cond_vib}
2 K \eta > \tilde{N} J_0(kx_m)^2 \;,
\end{equation}
where $\tilde{N} = \gamma \eta / (1+ \Omega^2 T_1^2)$ is the strength of nonlinear interactions \cite{Schwartz_PRL}, and
$J_0(kx_m)^2$ is the attenuation factor due to the crystal vibration. The similar condition for rotation sensing in the
case of a toroidal Bose-Einstein condensate, as derived in \cite{Leggett2}, is $V_0>g$ (where $V_0$ is the asymmetry
energy and $g$ the mean (repulsive) interaction energy per particle in the s-wave state). It becomes, in the presence of
Feshbach resonance induced by a magnetic field $B$ as described in \cite{Nature98}~:
\begin{equation} \label{cond_fesh}
V_0 > g \left( 1 - \frac{\Delta}{B-B_0} \right) \;,
\end{equation}
where $\Delta$ and $B_0$ are characteristic parameters. In this equation, $g$ represents the nonlinear interactions, and
the attenuation factor is $1-\Delta/(B-B_0)$. Condition (\ref{cond_fesh}) shows strong similarities with condition
(\ref{cond_vib}). In both cases, rotation sensing is favored if nonlinear interactions are lowered, the ideal case being
$B=B_0+\Delta$ for the toroidal Bose-Einstein condensed gas and $J_0(kx_m)=0$ for the solid-state ring laser. The
parameter for the control of nonlinear interactions is the magnetic field $B$ in the first case and the movement
amplitude $x_m$ in the second case.

In conclusion, we have developed a concrete method for tuning and suppressing nonlinear interactions in the case of a
solid-state ring laser, by vibrating the gain crystal along the cavity axis. Our theoretical model shows a very good
agreement with the experiment. The observation of rotation sensing in the solid-state ring laser allows the direct
measurement of the strength of nonlinear interactions, leading to the experimental demonstration of their fine tuning
and even suppression. Furthermore, following the previous work of \cite{Schwartz_PRL}, we have underlined the analogy
between our system and other ring-shaped macroscopic quantum configurations where nonlinear interactions could be tuned,
for example a Bose-Einstein condensed gas with magnetically-induced Feshbach resonance. This illustrates the richness of
such devices, both from applicative and fundamental perspectives.

The authors thank M. Defour, M. Mary, E. Bonneaudet and Thales Aerospace Division for constant support. They are also
grateful to A. Aspect and A. Mignot for fruitful discussions, and to F. Grabisch for his contribution to the numerical
simulations.

\end{document}